\documentclass[12pt,oneside]{artikle}
\usepackage{times}
\usepackage{mathptmx}
\usepackage{cite}
\usepackage{url}
\begin{document}

\begin{flushright}
LA-UR-07-8412
\end{flushright}

\title{$\pi\pi$ phase shifts from $K\rightarrow2\pi$}
\author{The FlaviaNet Kaon Working Group
\thanks{\texttt{http://www.lnf.infn.it/wg/vus}}
\thanks{The members of the FlaviaNet Kaon Working Group who contributed 
most to this note are 
V.~Cirigliano (Los Alamos); and 
C.~Gatti, M.~Moulson, and M.~Palutan (Frascati).}
}
\date{31 July 2008}
\maketitle

\begin{abstract}
We update the numerical results for the s-wave $\pi\pi$
scattering phase-shift difference $\delta_0^0-\delta_0^2$
at $s = m_K^2$ 
from a previous study of isospin breaking in $K\to2\pi$
amplitudes in chiral perturbation theory. We include recent data
for the $K_S\to\pi\pi$ and $K^+\to\pi^+\pi^0$ decay widths and include
experimental correlations.
\end{abstract}

The authors of Refs.~\citen{vincenzokpp1} and~\citen{vincenzokpp2} 
have shown the importance of taking into account radiative corrections 
in $K\rightarrow\pi\pi$ decays both in experimental measurements and 
theoretical predictions. In particular, these corrections are enhanced
in the extraction of the phase shifts $\delta_0^0-\delta_0^2$ by the 
$\Delta I=1/2$ rule by a factor of $\sim$22. On the experimental side, 
KLOE has measured with high accuracy the ratio of branching ratios (BRs) 
for the decay
 $K_{S}\rightarrow\pi^{+}\pi^{-}(\gamma)$
to the decay
 $K_{S}\rightarrow\pi^0\pi^0$ \cite{kloekspp}.
This measurement is fully inclusive of radiation for the 
$\pi^+\pi^-(\gamma)$ channel, allowing an unambiguous comparison 
with theoretical predictions.
For the extraction of the phase shift $\delta_0^0-\delta_0^2$ 
from $K\rightarrow\pi\pi$ we follow Ref.~\citen{vincenzokpp2}.
In the presence of electromagnetic interactions,
the usual isospin decomposition of amplitudes becomes:
\begin{eqnarray}
  \nonumber
      {\cal A}_{+-} &=& {\cal A}_{1/2}+\frac{1}{\sqrt{2}} 
        ({\cal A}_{3/2}+{\cal A}_{5/2})
      \\
      \nonumber
      {\cal A}_{00} &=& {\cal A}_{1/2}-\sqrt{2}({\cal A}_{3/2}+{\cal A}_{5/2})
      \\
      {\cal A}_{+0} &=& \frac{3}{2}({\cal A}_{3/2}-\frac{2}{3}{\cal A}_{5/2})
      \label{finiteampl}
\end{eqnarray}
where ${\cal A}_{+-,+0}$ are the infrared-finite amplitudes for the decays 
$K^{0}\rightarrow\pi^{+}\pi^{-}(\gamma)$
and $K^{+}\rightarrow\pi^{+}\pi^{0}(\gamma)$, respectively, 
and ${\cal A}_{00}$ is the amplitude for the decay 
$K^{0}\rightarrow\pi^{0}\pi^{0}$. These amplitudes are related 
to the decay rate by: 
\begin{equation}
  |{\cal A}_{n}| = 
    \left(\frac{2\sqrt{s_{n}}\Gamma_{n}}{G_{n}\Phi_{n}}\right)^{1/2}
  \label{expampl}
\end{equation}
where $n=\{+-,00,+0\}$. The factor $G_{n}$ is associated with the effect of 
real and virtual photons \cite{vincenzokpp1}. 
The following notation is introduced in 
Ref.~\citen{vincenzokpp2}:
\begin{eqnarray}
  \nonumber
  A_{0} e^{i\chi_{0}} &=& {\cal A}_{1/2}
  \\
  \nonumber
  A_{2} e^{i\chi_{2}}&=& {\cal A}_{3/2}+{\cal A}_{5/2}  
  \\
  A_{2}^{+} e^{i\chi_{2}^{+}}&=& {\cal A}_{3/2}-\frac{2}{3}{\cal A}_{5/2}  
  \label{isospinampl}
\end{eqnarray}
where in the absence of electromagnetic interactions the $A_{I}$ are the 
standard isospin amplitudes
and the phases $\chi_{I}$ are identified with the s-wave $\pi\pi$-scattering
phase shifts $\delta_0^I(s=m_K^2)$.
Otherwise, we have, by Eqs.~(7.20) and~(7.33) of Ref.~\citen{vincenzokpp2}:
\begin{equation}
  \delta_0^0-\delta_0^2 = \chi_{0}-\chi_{2}+(6.2\pm3.0)^{\circ}
  \label{deltacorr}
\end{equation}
and:
\begin{eqnarray}
  \nonumber
  |A_{0}|^{2}&=& a_{0} g_{8}^{2}+b_{0} g_{27}^{2}+c_{0} g_{8} g_{27}
  \\
  \nonumber
  |A_{2}|^{2}&=& a_{2} g_{8}^{2}+b_{2} g_{27}^{2}+c_{2} g_{8} g_{27}
  \\
  |A_{2}^{+}|^{2}&=& a_{2}^{+} g_{8}^{2}+b_{2}^{+} g_{27}^{2}+c_{2}^{+} g_{8} g_{27}
  \label{sviluppo}
\end{eqnarray}
where $g_{8,27}$ are the coefficients of the leading ($\mathcal{O}(p^2)$) octet 
and 27-plet weak non-leptonic chiral operators (as defined in Eqs.~(2.6)
and~(2.7) of Ref.~\citen{vincenzokpp2}). 
Note that in order to arrive at Eqs.~(\ref{sviluppo}), a number of 
higher-order chiral effective couplings are needed.
Here we adopt the estimates given in Section 5 of Ref.~\citen{vincenzokpp2}.

The coefficients $a$, $b$, and $c$ depend on the chiral
renormalization scale $\nu_{\chi}$ (from the matching uncertainty in the 
low-energy constants)
and the tree-level $\pi^{0}-\eta$ mixing angle $\varepsilon^{(2)}$ given by:
\begin{equation}
  \varepsilon^{(2)} = \frac{\sqrt{3}}{2} \frac{m_{d}-m_{u}}{2 m_{s}-m_{d}-m_{u}}.
\end{equation} 
We have obtained the values of these coefficients
from the authors of  
Ref.~\citen{vincenzokpp2}.

Eqs.~(\ref{finiteampl}), (\ref{expampl}), and~(\ref{isospinampl}) can 
be combined to obtain
\begin{eqnarray}
  \nonumber
  A_{2}^{+}&=&\frac{2}{3}   |{\cal A}_{+0}| 
  \\
  \nonumber
  (A_{0})^{2}+(A_{2})^{2}&=&
  \frac{2}{3}|{\cal A}_{+-}|^{2}+\frac{1}{3}|{\cal A}_{00}|^{2}
  \\
  \frac{A_{2}}{A_{0}}\cos{(\chi_{0}-\chi_{2})}&=&
       \frac{r-1+(\frac{A_{2}}{A_{0}})^{2}(2r-\frac{1}{2})}{\sqrt{2}(1+2r)}
  \label{chi02eq}
\end{eqnarray}
where $r=|{\cal A}_{+-}/{\cal A}_{00}|^{2}$. 
Using the expansion of Eq.~(\ref{sviluppo}),
this system of equations can be written in terms of the three 
unknowns, $\chi_{0}-\chi_{2}$, $g_{8}$, and $g_{27}$.
We solve this system using a numerical minimization procedure. 

The experimental inputs used to obtain the amplitudes by Eq.~(\ref{expampl}) 
are listed in Table~\ref{tab:exp}. As noted above, the BRs for 
$K_S\to\pi^+\pi^-(\gamma)$ and $K_S\to\pi^0\pi^0$ are obtained 
from the KLOE measurement of their ratio, which accounts for the 
large value of the correlation coefficient.
The BR for $K^+\to\pi^+\pi^0$ is weakly correlated with the value of the 
$K^+$ lifetime by the fit performed in Ref.~\citen{Flavia}.
These correlations are
taken into account in the minimization procedure.
While the KLOE measurement of the ratio of $K_S\to\pi\pi$ BRs is fully 
inclusive of radiation in the $\pi^+\pi^-(\gamma)$ channel, the
inclusiveness of the value for ${\rm BR}(K^+\to\pi^+\pi^0)$ 
from the fit in Ref.~\citen{Flavia} is less well defined. However,
this is of lesser concern because
of the dominance of the $I=1/2$ amplitudes. 
\begin{table}
\begin{center}
\begin{tabular}{lccc}
\hline\hline
Parameter & Value & Correlation & Reference \\
\hline
$\begin{array}{l}
{\rm BR}(K_S\to\pi^+\pi^-(\gamma)) \\
{\rm BR}(K_S\to\pi^0\pi^0)
\end{array}$ &
$\begin{array}{l}
\mbox{0.69196(51)} \\
\mbox{0.30687(51)} 
\end{array}$ &
$-\mbox{0.9996}$ &
\cite{kloekspp} \\
\hline
$\begin{array}{l}\tau_S\end{array}$ & 
$\begin{array}{c}\mbox{0.08958(5)~ns}\end{array}$ & & \cite{PDG} \\
\hline
$\begin{array}{l}
{\rm BR}(K^+\to\pi^+\pi^0) \\
\tau_+
\end{array}$ &
$\begin{array}{c}
\mbox{0.2064(8)} \\
\mbox{12.384(19)~ns}
\end{array}$ &
$-\mbox{0.032}$ &
\cite{Flavia} \\
\hline\hline
\end{tabular}
\end{center}
\caption{Experimental inputs used to obtain $g_8$, $g_{27}$, and $\chi_0 - \chi_2$.}
\label{tab:exp}
\end{table}
With $\nu_{\chi}=\mbox{0.77}$~GeV and
$\varepsilon^{(2)}=1.06\times10^{-2}$ we obtain:
\begin{eqnarray}
  \nonumber
  g_{8} &=& 3.6435\pm0.0010
  \\
  \nonumber
  g_{27} &=& 0.2987\pm0.0006
  \\
  \chi_{0}-\chi_{2} &=& (51.26\pm0.82)^{\circ}
  \label{chi02result}
\end{eqnarray}
The error matrix is:
\begin{equation}
  \left(
  \begin{array}{ccc}
    1.08\times10^{-6} & -7.2\times10^{-8} & -2.7\times10^{-5} \\
    $-$ & 3.8\times10^{-7} & 5.2\times10^{-5} \\
    $-$ & $-$ & 0.68 \\
  \end{array}
  \right)
  \label{statmatphase}
\end{equation}

We compute the systematic error by varying the chiral renormalization scale
parameter $\nu_{\chi}$
from 0.5 GeV to 1~GeV, 
and the tree-level $\pi^{0}-\eta$ mixing angle 
$\varepsilon^{(2)}$ from $0.6\times10^{-2}$ to $1.5\times10^{-2}$, and taking
half of the total variation as the uncertainty.
The systematic error matrix is:
\begin{equation}
  \left(
  \begin{array}{ccc}
    4.0\times10^{-2} & 3.8\times10^{-4} & -0.28  \\
    $-$  & 9.3\times10^{-6} & -2.7\times10^{-3}  \\
    $-$  & $-$  & 2.0 \\
  \end{array}
  \right)
  \label{systmatphase}
\end{equation}

Including the systematic errors and using Eq.~(\ref{deltacorr}) we have:
\begin{eqnarray}
  \nonumber
  g_{8} &=& (3.644\pm0.001_{\rm exp}\pm0.200_{\nu_{\chi}\oplus\varepsilon^{(2)}})
       =(3.64\pm0.20)
  \\
  \nonumber
  g_{27} &=& (0.2987\pm0.0001_{\rm exp}\pm0.0030_{\nu_{\chi}\oplus\varepsilon^{(2)}})
       =(0.2987\pm0.0030)
  \\
  \delta_0^0-\delta_0^2 &=& (57.5\pm0.8_{\rm exp}\pm 3.0_{\gamma_2}
           \pm 1.4_{\nu_{\chi}\oplus\varepsilon^{(2)}})^{\circ}
       = (57.5\pm3.4)^{\circ} 
    \label{phaseresult}
\end{eqnarray}

The results in Eqs.~(\ref{phaseresult}) have been obtained using 
the central values for the higher-order chiral couplings estimated 
in Ref.~\citen{vincenzokpp2} within the large-$N_C$ expansion.
The quoted uncertainty reflects only the uncertainty in the matching scale.  
As already pointed out in Ref.~\citen{vincenzokpp2} (Section 7.3),  
the extraction of $g_8$ and $g_{27}$ is rather sensitive to the input on the 
chiral couplings. For instance, even a simple variant of the large-$N_C$ 
procedure leads to changes in $g_8$ and $g_{27}$ at the $10\%$ 
level~\cite{vincenzokpp2}. This implies that the values for $g_{8,27}$ 
quoted in Eq.~(\ref{phaseresult}) are affected by an unknown systematic 
offset of at least $10\%$. (See also the analysis of Ref.~\citen{BB05}.)
On the other hand, the extraction of the
phase difference $\chi_0 - \chi_2$ is quite insensitive to the input on 
the effective chiral couplings. The result for the phase difference
$\delta_0^0-\delta_0^2$ instead depends quite sensitively on the estimate
of the isospin-breaking correction of Eq.~(\ref{deltacorr}).

It is interesting to compare the present result for $\delta_0^0-\delta_0^2$
with predictions from phenomenological evaluations. 
The Roy equations \cite{Roy71} determine the $\pi\pi$ scattering
amplitude in terms of its imaginary part at intermediate energies,
up to two subtraction constants: the s-wave scattering lengths
$a_0^0$ and $a_0^2$.
Colangelo, Gasser, and Leutwyler \cite{CGL01} obtain values for 
$a_0^0$ and $a_0^2$ by matching a representation of the $\pi\pi$
scattering amplitude from $\mathcal{O}(p^6)$ calculations in chiral 
perturbation theory with a phenomenological representation 
based on the Roy equations. They obtain
$\delta_0^0 - \delta_0^2 = (47.7\pm1.5)^\circ$, which differs from our
result by $2.6\sigma$.
Kami\'nski, Pel\'aez, and Yndur\'ain \cite{KPY08} fit 
experimental $\pi\pi$ scattering amplitudes at both low and high 
energies with parameterizations that satisfy analyticity at low
energy, constrained to satisfy the forward 
dispersion relations and the Roy equations. They obtain 
$\delta_0^0 - \delta_0^2 = (50.9\pm1.2)^\circ$, which differs from
our result by $1.8\sigma$. It is important
to note that their input data 
includes low-energy s-wave phase shift determinations from 
$K_{e4}$ and $K\to2\pi$ decays, including a preliminary result
for $\delta_0^0 - \delta_0^2$ from this update, differing very little
from the value presented here.
The significant discrepancies between our result, based on
$K_S$ and $K^+$ BR measurements, and the results of phenomenological
analyses of $\pi\pi$ scattering (with or without constraints from
chiral symmetry) are puzzling and deserve further investigation.

\section*{Acknowledgements}
We thank all members of the FlaviaNet Kaon Working Group
(\url{http://www.lnf.infn.it/wg/vus}) for comments, discussion,
and suggestions. This work is supported in part by EU Contract
MTRN-CT-2006-035482 (FlaviaNet).

\end{document}